\documentclass[usenatbib]{mn2e}
\usepackage{graphicx,natbib,times,amssymb}
\def\dsm{$M_\odot$}
\def\dsr{$R_\odot$}
\def\dsl{$L_\odot$}
\def\teff{$T_{\rm eff}$}
\def\dov{$\delta_{\rm ov}$}
\def\dra{$r_{01}$}
\def\drb{$r_{10}$}

\def\dhz{$\mu$Hz}

\title[ASTEROSEISMIC ANALYSIS OF KEPLER TARGET KIC 2837475]
{ASTEROSEISMIC ANALYSIS OF KEPLER TARGET KIC 2837475}
\author[Yang et al.]{Wuming Yang$^{1,2}$\thanks{E-mail:yangwuming@bnu.edu.cn},
Zhijia Tian$^{1}$, Shaolan Bi$^{1}$, Zhishuai Ge$^{1}$, Yaqian Wu$^{1}$, Jinghua Zhang$^{1}$ \\
$^{1}$Department of Astronomy, Beijing Normal University, Beijing 100875,
China.\\
$^{2}$School of Physics and Chemistry, Henan Polytechnic University,
Jiaozuo 454000, Henan, China. \\}

\begin{document}

\date{ }

\pagerange{\pageref{firstpage}--\pageref{lastpage}} \pubyear{2010}

\maketitle

\label{firstpage}

\begin{abstract}
The ratios $r_{01}$ and $r_{10}$ of small to large separations of KIC 2837475 primarily
exhibit an increase behavior in the observed frequency range. The calculations indicate
that only the models with overshooting parameter $\delta_{\rm ov}$ between approximately
1.2 and 1.6 can reproduce the observed ratios $r_{01}$ and $r_{10}$ of KIC 2837475.
The ratios $r_{01}$ and $r_{10}$ of the frequency separations of p-modes with inner turning
points that are located in the overshooting region of convective core can exhibit an increase behavior.
The frequencies of the modes that can reach the overshooting region decrease with the increase
in $\delta_{\rm ov}$. Thus the ratio distributions are more sensitive to $\delta_{\rm ov}$
than to other parameters. The increase behavior of the KIC 2837475 ratios results from
a direct effect of the overshooting of convective core. The characteristic of the ratios
provides a strict constraint on stellar models. Observational constraints point to a star with
$M=1.490\pm0.018$ $M_{\odot}$, $R=1.67\pm0.01$ $R_{\odot}$, age $=2.8\pm0.4$ Gyr,
and $1.2\lesssim \delta_{\rm ov} \lesssim1.6$ for KIC 2837475.
\end{abstract}


\section{Introduction}
By comparing observed oscillation frequencies and the ratios of
small to large separations with those calculated from theoretical
models, asteroseismology imposes strict constraints on stellar models.
The small separations are defined as \citep{roxb03}
\begin{equation}
d_{10}(n)\equiv-\frac{1}{2}(-\nu_{n,0}+2\nu_{n,1}-\nu_{n+1,0})
\label{d10}
\end{equation}
and
\begin{equation}
d_{01}(n)\equiv\frac{1}{2}(-\nu_{n,1}+2\nu_{n,0}-\nu_{n-1,1}).
\label{d01}
\end{equation}
In calculation, equations (\ref{d10})
and (\ref{d01}) are generally rewritten as the smoother five-point
separations.

Asteroseismology has proved to be a powerful
tool for determining fundamental star parameters,
diagnosing internal structures of stars,
and probing physical processes in stellar interiors
\citep{egge05, yang07b, stel09, chri10, yang10, yang11a,
yang12, silv11, silv13, liu14, chap14, metc14, guen14}.

F-type main-sequence (MS) stars usually have a convective core
and a convective envelope. Consequently, they have various
characteristics that are similar to that of the Sun,
such as oscillations. On the other hand, they have peculiar
properties due to the convective core that leads to a large
gradient of chemical compositions at the bottom of
the radiative region and an overshooting of the core
in their interior. The overshooting of convective core
extends the region of chemical mixing by a distance \dov{}$Hp$
above the top of the convective core that is determined
by Schwarzchild criterion, where $Hp$ is the local pressure
scale height and \dov{} is a free parameter. Consequently,
the overshooting of the core brings more H-rich material
into the core, which prolongs the lifetime of the burning
of core hydrogen, enhances the He-core mass left behind,
and strongly changes the global characteristics of
the following giant stages \citep{schr97, yang12},
such as the critical mass of He-flash and the global
oscillation properties of red-clump stars \citep{yang12}.

The parameter \dov{} has been observationally estimated in
several ways. \cite{prat74} and \cite{dema94} found that
the value of \dov{} is approximately 0.1-0.2 by
comparing the theoretical and observational color-magnitude
diagram of clusters. In addition, the value of \dov{}
estimated by matching the exact properties of certain
$\zeta$ Aurigae eclipsing binaries is roughly 0.2-0.3 \citep{schr97}.
The uncertainty of the mass and extension of the convective
core due to overshooting directly affects the determination
of the global parameters of stars by asteroseismology or
other studies based on stellar evolution \citep{mazu06}.
Thus determining the presence of the convective core and its
extension is important for understanding the structure and
evolution of stars.

The p-modes with $l=1$ penetrate more deeply into stellar
interiors than the higher-$l$ modes. Thus, the ratios $r_{01}$
and $r_{10}$ are a potential tool for probing the existence
and extension of a convective core of stars.
Many authors \citep{mazu06, cunh07, dehe10, deme10,
silv11, silv13, liu14, guen14, tian14} have studied the
overshooting of convective cores by asteroseismology.
\cite{tian14} gave that the value of \dov{} is in the range
of 0.0-0.2 for KIC 6225718. \cite{liu14} found
that it is in the range of 0.4 and 0.8 for HD 49933;
however, \cite{guen14} obtained the \dov{} for Procyon
in the range of 0.9 and 1.5, which is much larger than
the generally accepted value. This large \dov{} may exist in
other stars, and if this large \dov{} is confirmed in more stars,
the general understanding of the structure and evolution
of stars will be improved to a certain degree.

Moreover, although the increase behaviors in the ratios
of KIC 6106415 and KIC 12009504 arise from the low
signal-to-noise ratio and the larger linewidth at the
high-frequency end \citep{silv13}, the increase in the
ratios of HD 49933 at high frequencies may result from
the effects of overshooting of convective core \citep{liu14}.
\cite{liu14} argued that the gradient of the mean molecule weight
in the radiative region hinders the propagation of p-modes, while
the hindrance does not exist in the convective core. Therefore, the
ratios \dra{} and \drb{} can exhibit an increase with frequencies
when the inner turning points of the corresponding modes with $l=1$ are
located in the overshooting region.

Individual frequencies of p-modes of dozens of MS stars
have been extracted by \cite{appo12}, which provides an
opportunity for studying the overshooting of the convective core and
the behaviors of ratios \dra{} and \drb{}. In these stars, the
ratios \dra{} and \drb{} of KIC 2837475 increase linearly with a
frequency in the range of approximately 1050 and 1600 \dhz{},
which may be related to the overshooting
of convective core. The mass of KIC 2837475 estimated by \cite{chap14} is
$1.41^{+0.06}_{-0.04}$ \dsm{} for the constraints of the spectroscopic
\teff{} and [Fe/H] of \cite{brun12} or $1.47^{+0.15}_{-0.13}$ \dsm{}
for the constraints of the IRFM \teff{} and field-average [Fe/H].
Moreover, \cite{metc14} determined the value to be $1.39\pm0.06$ \dsm{}
for KIC 2837475.

In this work, we focus mainly on whether the observed characteristics of
KIC 2837475 can be directly reproduced by stellar models. In order
to find the best model of KIC 2837475, we use the chi squared method. Firstly,
using the constraints of luminosity, \teff{}, and [Fe/H], we obtain an
approximate set of solutions. Then around this set of solutions, we seek for
best models that match both non-seismic constraints and the individual
frequencies extracted by \cite{appo12}. And finally, we compare
\dra{} and \drb{} of models with the observed \dra{} and \drb{}.
In Section 2, the stellar models are introduced.
In Section 3, the observational constraints and the calculated results
are presented, and in Section 4, the results are summarized and discussed.

\section{STELLAR MODELS}
\begin{figure}
\centering
\includegraphics[scale=0.5, angle=-90]{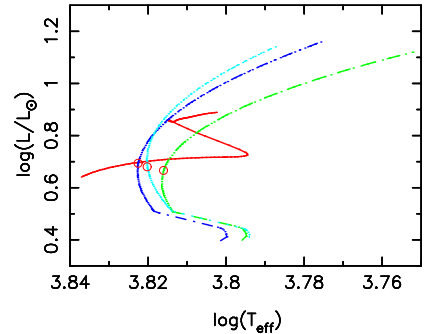}
\caption{The evolutionary tracks of models with $M=1.50$ \dsm{}
with different \dov{} and $\alpha$. The solid (red) line
represents the model track with \dov{} = 0.0; the dashed (green)
line represents \dov{} = 1.2; the dash-dotted (blue) line
represents \dov{} = 1.4; the dotted (cyan) line represents \dov{} = 1.6.
The circles indicate the most likely models for KIC 2837475. }
\label{fhd1}
\end{figure}

To study the solar-like oscillations of KIC 2837475, a grid of
evolutionary tracks was computed using the Yale Rotation Evolution
Code \citep{pins89, yang07a} in its non-rotation configurations.
The OPAL EOS tables \citep{roge02}
and OPAL opacity tables \citep{igle96} were used, supplemented by
the \cite{alex94} opacity tables at low temperature.
Convection is treated according to the standard mixing-length theory.
The value of the mixing-length parameter $\alpha$ calibrated to
the Sun is 1.74; in this work, it is a free parameter.
The overshooting of the convective core is
described by the parameter \dov{}. The full mixing of chemical
compositions is assumed in the overshooting region in our models.
The diffusion and settling of both helium and heavy elements are computed
by using the diffusion coefficients of \cite{thou94} for models with a mass
less than 1.30 \dsm{}. The range of the values of parameters,
mass ($M$), $\alpha$ and \dov{}, and the chemical compositions of zero-age
MS models are summarized in Table \ref{tab1}, where the $\delta$ indicates
the resolution of the parameters. All models are computed from zero-age MS
to the end of MS or the subgiant stage. For example, Figure \ref{fhd1} shows
the evolutionary tracks of models with $M=1.50$ \dsm{} with varying
\dov{} and $\alpha$.

\begin{table}
\begin{center}
\caption[]{The Range of the Parameters of Zero-age MS Models.}
\label{tab1}
\begin{tabular}{ccccc}
  \hline\noalign{\smallskip}
  \hline\noalign{\smallskip}
   Variable    & Minimum  & Maximum & $\delta$  \\
   \hline\noalign{\smallskip}
       M/M$_\odot$ & 1.00     & 1.60   & 0.02    \\
      $\alpha$    & 1.65     & 2.05 &  0.1     \\
      $\delta_{\rm ov}$ &  0.0     & 1.8 &  0.2     \\
      $Z_{i}$  & 0.010    & 0.050 & 0.002   \\
      $ X_{i} $       & 0.655    & 0.742  & 0.002   \\
   \hline\noalign{\smallskip}

  \noalign{\smallskip}
\end{tabular}
\end{center}
\end{table}

The adiabatic frequencies of the low-degree p-modes of models
are computed by using the Guenther pulsation code \citep{guen94}.
For the modes with a given degree $l$, the frequencies
$\nu_{corr}(n)$ corrected from the near-surface effects of a model
are calculated by using equation \citep{kjel08}
\begin{equation}
\nu_{corr}(n)=\nu_{mod}(n)+a[\frac{\nu_{mod}(n)}{\nu_{max}}]^{b},
\label{nucor}
\end{equation}
where $b$ is fixed to be 4.9, $a$ is determined from the observed
frequencies of the modes with the given $l$ and the adiabatic frequencies
of the modes with the degree $l$ of the model by using equations
(6) and (10) of \cite{kjel08},
$\nu_{mod}(n)$ represents the adiabatic frequencies of the model,
$\nu_{max}$ is the observed frequency of maximum oscillation
power\footnote{The corrected and uncorrected adiabatic frequencies
of models used here can be downloaded from
http://pan.baidu.com/s/1c00LPwk}.

\section{Observational Constraints}

\subsection{Non-asteroseismic Observational Constraints}
KIC 2837475 is an F5 star \citep{wright03}. Its effective
temperatures are $6562^{+68}_{-75}$ K \citep{ammo06},
$6740\pm70$ K \citep{brun12}, or $6462\pm125$ K \citep{mol13}.
The value of [Fe/H] for KIC 2837475 is $0.21\pm0.12$ \citep{ammo06},
$-0.02\pm0.06$ \citep{brun12}, or $-0.06\pm0.21$ \citep{mol13}.
Combining the value of $(Z/X)_{\odot}=0.023$ given by \cite{grev98},
the ratio of surface heavy-element abundance to hydrogen abundance,
$(Z/X)_{s}$, can be obtained and is approximately between 0.028
and 0.049 for the [Fe/H] of \cite{ammo06}, between 0.019 and 0.025 for the
[Fe/H] of \cite{brun12}, or between 0.012 and 0.033 for the [Fe/H] of
\cite{mol13}. Moreover, its visual magnitude is $8.547\pm0.013$ mag \citep{ammo06}.
The bolometric correction can be estimated from the tables of \cite{flow96};
while extinction is obtained from \cite{ammo06}.
By combining its parallax $\Pi=10.3^{+5.8}_{-5.0}$ mas \citep{ammo06},
its luminosity is estimated as approximately $6.0\pm4.8$ \dsl{}.

For each model, the value of $\chi_{c}^{2}$ is calculated.
The function $\chi_{c}^{2}$ is defined as
\begin{equation}
\chi^{2}_{c} = \frac{1}{3}\sum_{i=1}^{3}
[\frac{C_{i}^{theo}-C_{i}^{obs}}{\sigma(C_{i}^{obs})}]^{2},
\end{equation}
where the quantities $C_{i}^{obs}$ and $C_{i}^{theo}$ are the observed
and model values of \teff{}, $L/L_{\odot}$, and $(Z/X)_{s}$,
respectively. The observational uncertainty is indicated
by $\sigma(C_{i}^{obs})$. In the first step, the models
with $\chi_{c}^{2}\leq1$ are used as candidates for KIC 2837475.

\subsection{Asteroseismic Constraints}
To find the models that can reproduce the properties
of KIC 2837475 , the value of $\chi_{\nu}^{2}$ was computed.
The function $\chi_{\nu}^{2}$ is defined as
\begin{equation}
\chi_{\nu}^{2} = \frac{1}{N}\sum_{i=1}^{N}
[\frac{\nu_{i}^{theo}-\nu_{i}^{obs}}{\sigma(\nu_{i}^{obs})}]^{2},
\end{equation}
where $\nu_{i}^{obs}$ and $\nu_{i}^{theo}$ are the observed and
corresponding model eigenfrequencies of the $i$th mode, respectively,
and $\mathbf{\sigma(\nu_{i}^{obs})}$ is the observational uncertainty of
the $i$th mode. The value of the $N$ is 44 for KIC 2837475. We
also calculated the value of $\chi_{\nu_{corr}}^{2}$ of models.
The models with $\chi_{c}^{2}\leq1.0$ and $\chi_{\nu}^{2}$ or
$\chi_{\nu_{corr}}^{2}\lesssim10.0$ are chosen as candidates.

\begin{figure}
\centering
\includegraphics[scale=0.35, angle=-90]{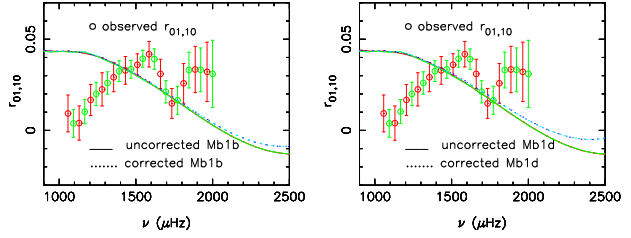}
\caption{The distributions of ratios \dra{} and \drb{} of models Mb1b and Mb1d
as a function of frequency. The uncorrected ratios are calculated from adiabatic
frequencies $\nu_{n,l}$ of the models, while the corrected ratios are computed from
the corrected frequencies $\nu_{corr}$ of the models.}
\label{fpgba}
\end{figure}

In order to ensure finding the model with the minimum $\chi_{\nu}^{2}$
and the model with the minimum $\chi_{\nu_{corr}}^{2}$, the time-step of
the evolution for each track is set as small as possible when the model
evolves to the vicinity of the error-box of luminosity and effective
temperature in the H-R diagram. This makes the consecutive models have
an approximately equal $\chi_{\nu}^{2}$ or $\chi_{\nu_{corr}}^{2}$.
The value of $\chi_{\nu_{corr}}^{2}$ of the model with the minimum $\chi_{\nu}^{2}$
is usually not a minimum (see the models Mb1d and Mm2b in Table \ref{tab2}).
In some cases, for the consecutive models in a track, the value of
the minimum $\chi_{\nu_{corr}}^{2}$ is even larger than
that of the minimum $\chi_{\nu}^{2}$ (see the models Mm2b and Mm2d
in Table \ref{tab2}). The difference between the age of the model with the minimum
$\chi_{\nu}^{2}$ and the age of the model with the minimum $\chi_{\nu_{corr}}^{2}$
is generally several Myr. However, the distributions of ratios \dra{} and \drb{}
of the two models are almost the same (see Figure \ref{fpgba}), i.e., the interior
structures of the two models are almost the same. For a given mass and \dov{},
the model with the minimum $\chi_{\nu_{corr}}^{2}$ was chosen as the best model.

\begin{figure}
\centering
\includegraphics[scale=0.35, angle=-90]{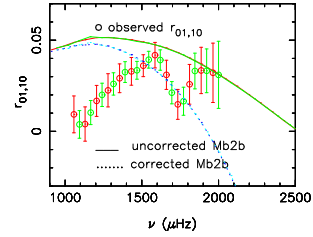}
\caption{The distributions of corrected and uncorrected ratios \dra{} and \drb{}
of model Mb2b. }
\label{fpgbb}
\end{figure}

Figures \ref{fpgba} and \ref{fpgbb} show that the distributions of
ratios \dra{} and \drb{} are affected by correction. \cite{kjel08} argued
that the offset from incorrect modeling of the near-surface layers is independent
of $l$. Thus we computed the value of $a$ in equation (\ref{nucor}) only from
frequencies of radial modes ($l=0$) and then apply it to all modes.
In this case, the distributions of ratios \dra{} and \drb{} are not affected
by the correction, i.e., the corrected and uncorrected ratios are the same;
however, the value of $\chi_{\nu_{corr}}^{2}$ becomes larger (see the value
in the parentheses in Table \ref{tab2}). In present work, the $a$ for the modes
with degree $l$ is calculated from the observed and model frequencies of the modes
with the degree $l$.

\subsubsection{The Models with the Effective Temperature and [Fe/H] of Bruntt}
For a given mass, the model minimizing $\chi_{\nu_{corr}}^{2}+\chi_{c}^{2}$
is chosen as a candidate for the best-fit model. The fundamental parameters of
the models are listed in Table \ref{tab2}, where some consecutive models
are listed to state that the value of $\chi_{\nu}^{2}$ of the model with the
minimum $\chi_{\nu_{corr}}^{2}$ is not a minimum. The model Mb1b
has the minimum $\chi_{\nu_{corr}}^{2}$ and $\chi_{\nu_{corr}}^{2}+\chi_{c}^{2}$,
suggesting that Mb1b are the best-fit model.

\begin{table*}
\begin{center}
\renewcommand\arraystretch{1.0}
\caption[]{Parameters of Models. The symbol $X_{c}$ indicates
the central hydrogen abundance of models. The symbol $r_{cc}$ shows
the radius of a convective core of models determined by Schwarzchild criterion;
while the $\chi_{\nu_{corr}}^{2}$ represents the chi squared of frequencies
corrected from near-surface effects. Models Mb's have the Effective temperature
and [Fe/H] of \cite{brun12}; while models Mm1 to Mm11 have those determined by \cite{mol13}.}
\label{tab2}
\begin{tabular}{p{0.70cm}cccccccccccccc}
  \hline\hline\noalign{\smallskip}
   Model & $M$ & \teff{}& $ L$ & $R$& age & $Z_{i}$ & $(Z/X)_{s}$ & $\alpha$
   & \dov{} & $X_{c}$  & $r_{cc}$ & $\chi_{\nu}^{2}$ & $\chi_{\nu_{corr}}^{2}$ & $\chi_{c}^{2}$ \\

  & (\dsm{}) & (K) & (\dsl{})  & (\dsr{}) & (Gyr) & &  & & & & (\dsr{}) & & \\
  \hline\hline\noalign{\smallskip}
 Mb1a & 1.30 & 6821 & 4.90 & 1.589 & 2.036 & 0.014 & 0.020 & 1.95 & 0.2 & 0.380 & 0.125 & 10.6 &  5.3 (6.1)$^{a}$ &  0.6 \\
 Mb1b & 1.30 & 6820 & 4.90 & 1.589 & 2.037 & 0.014 & 0.020 & 1.95 & 0.2 & 0.380 & 0.125 &  9.0 &  5.1 (6.0) &  0.6 \\
 Mb1c & 1.30 & 6820 & 4.90 & 1.589 & 2.038 & 0.014 & 0.020 & 1.95 & 0.2 & 0.379 & 0.125 &  8.0 &  5.4 (6.5) &  0.6 \\
 Mb1d & 1.30 & 6820 & 4.90 & 1.589 & 2.039 & 0.014 & 0.020 & 1.95 & 0.2 & 0.379 & 0.125 &  7.6 &  6.5 (7.6) &  0.6 \\
 Mb1e & 1.30 & 6819 & 4.90 & 1.589 & 2.040 & 0.014 & 0.020 & 1.95 & 0.2 & 0.379 & 0.125 &  7.8 &  8.1 (9.3) &  0.6 \\
 Mb2a & 1.32 & 6786 & 4.87 & 1.600 & 1.717 & 0.016 & 0.024 & 1.95 & 0.0 & 0.272 & 0.111 &  6.8 &  7.0 (6.2) &  0.3 \\
 Mb2b & 1.32 & 6785 & 4.87 & 1.600 & 1.721 & 0.016 & 0.024 & 1.95 & 0.0 & 0.271 & 0.111 &  6.5 &  6.3 (6.5) &  0.2 \\
 Mb2c & 1.32 & 6783 & 4.88 & 1.600 & 1.724 & 0.016 & 0.024 & 1.95 & 0.0 & 0.270 & 0.111 &  9.2 &  9.0 (12.9) &  0.2 \\
 Mb3 & 1.34 & 6772 & 4.89 & 1.611 & 1.851 & 0.018 & 0.026 & 1.95 & 0.2 & 0.400 & 0.131 &  5.8 &  5.2 (5.7) &  0.8 \\
  \hline\hline
 Mm1 & 1.19 & 6381 & 3.66 & 1.567 & 3.200 & 0.014 & 0.012 & 1.85 & 0.2 & 0.294 & 0.111 &  7.1 &  7.1 &  0.2 \\
 Mm2a & 1.21 & 6529 & 4.03 & 1.572 & 2.553 & 0.012 & 0.011 & 2.05 & 0.0 & 0.145 & 0.095 &  8.8 & 15.9 &  0.2 \\
 Mm2b & 1.21 & 6529 & 4.03 & 1.572 & 2.555 & 0.012 & 0.011 & 2.05 & 0.0 & 0.145 & 0.095 &  7.8 & 11.6 &  0.2 \\
 Mm2c & 1.21 & 6528 & 4.03 & 1.572 & 2.557 & 0.012 & 0.011 & 2.05 & 0.0 & 0.144 & 0.095 &  8.4 &  9.3 &  0.1 \\
 Mm2d & 1.21 & 6528 & 4.03 & 1.573 & 2.558 & 0.012 & 0.011 & 2.05 & 0.0 & 0.144 & 0.093 &  9.6 &  8.9 &  0.1 \\
 Mm2e & 1.21 & 6528 & 4.04 & 1.573 & 2.559 & 0.012 & 0.011 & 2.05 & 0.0 & 0.143 & 0.093 & 12.8 &  9.4 &  0.1 \\
 Mm3 & 1.23 & 6435 & 3.87 & 1.585 & 2.926 & 0.016 & 0.015 & 1.95 & 0.2 & 0.311 & 0.117 & 6.8 & 6.7 & 0.1\\
 Mm4 & 1.25 & 6392 & 3.80 & 1.592 & 2.793 & 0.018 & 0.016 & 1.85 & 0.2 & 0.334 & 0.118 & 7.5 &  6.9 &  0.2 \\
 Mm5 & 1.27 & 6479 & 4.07 & 1.603 & 2.719 & 0.018 & 0.018 & 2.05 & 0.2 & 0.325 & 0.112 & 7.7 &  7.1 &  0.1 \\
 Mm6 & 1.29 & 6424 & 3.98 & 1.611 & 2.662 & 0.020 & 0.020 & 1.95 & 0.2 & 0.341 & 0.123 & 7.5  & 6.9 &  0.1 \\
 Mm7 & 1.32 & 6657 & 4.52 & 1.600 & 1.753 & 0.018 & 0.027 & 1.75 & 0.0 & 0.293 & 0.113 & 8.6 & 6.4 &  0.9 \\
 Mm8 & 1.34 & 6528 & 4.29 & 1.622 & 2.142 & 0.022 & 0.032 & 1.75 & 0.2 & 0.392 & 0.129 & 9.0 & 7.4 & 0.5 \\
 Mm9 & 1.36 & 6588 & 4.46 & 1.622 & 1.801 & 0.020 & 0.029 & 1.75 & 0.0 & 0.311 & 0.116 & 7.1 & 6.8 &  0.5 \\
 Mm10 & 1.38 & 6589 & 4.50 & 1.629 & 1.800 & 0.024 & 0.035 & 1.75 & 0.2 & 0.427 & 0.135 &  6.2 &  4.7 &  1.0 \\
 Mm11 & 1.40 & 6583 & 4.54 & 1.641 & 1.617 & 0.024 & 0.035 & 1.75 & 0.0 & 0.344 & 0.122 &  6.7 &  6.4 &  0.9 \\
 \noalign{\smallskip}\hline\hline
\end{tabular}
\end{center}
 $^{a}$ The corrected frequencies of modes with $l=1$ and $2$ are calculated
 by using the $a$ that is determined from the frequencies of modes with $l=0$.
\end{table*}

\begin{figure}
\centering
\includegraphics[scale=0.35, angle=-90]{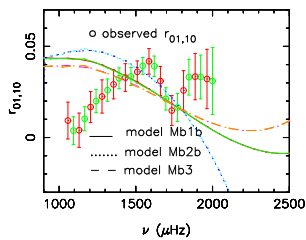}
\caption{The distributions of the observed and theoretical ratios \dra{} and \drb{}
as a function of frequency. The theoretical ratios are calculated from the corrected
frequencies $\nu_{corr}$ of models.}
\label{fpgb1}
\end{figure}

However, Figure \ref{fpgb1} shows that these models cannot reproduce
the observed ratios \dra{} and \drb{}. This hints us that the internal
structures of these models may be not consistent with that of KIC 2837475.

The effective temperatures of KIC 2837475 determined by \cite{mol13}
and \cite{ammo06} are obviously lower than that estimated by \cite{brun12}.
Our models with the effective temperature and [Fe/H] of \cite{brun12} cannot
reproduce the observed ratios \dra{} and \drb{}. Thus the models with the \teff{}
and [Fe/H] determined by \cite{mol13} and \cite{ammo06} also should be considered.

\subsubsection{The Models with the Effective Temperature and [Fe/H] of Molenda-Zakowicz}

\begin{figure}
\centering
\includegraphics[scale=0.35, angle=-90]{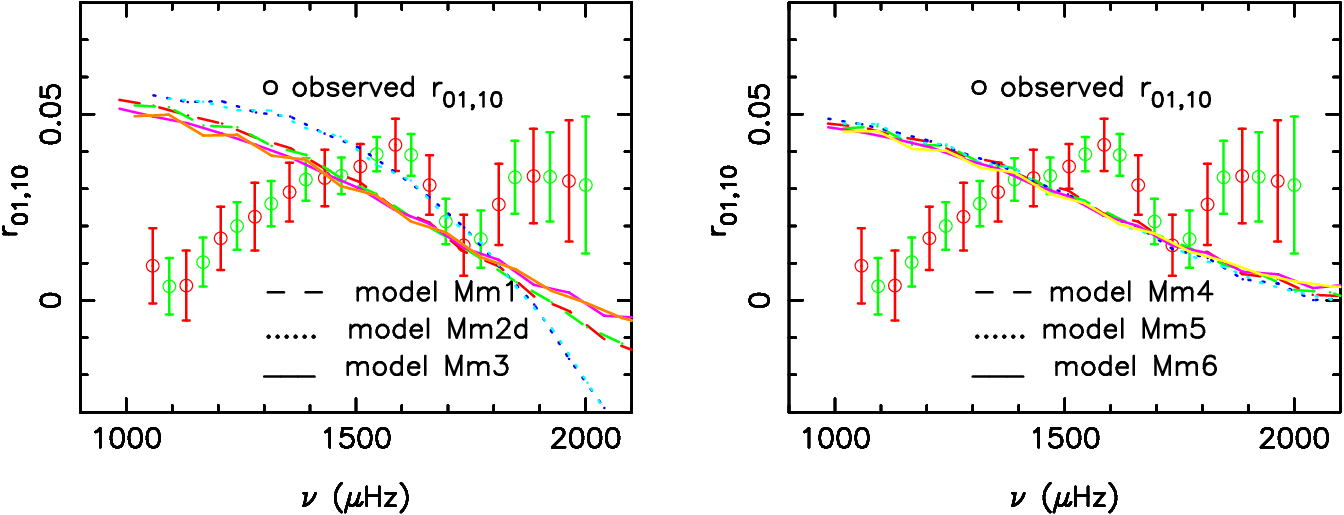}
\includegraphics[scale=0.35, angle=-90]{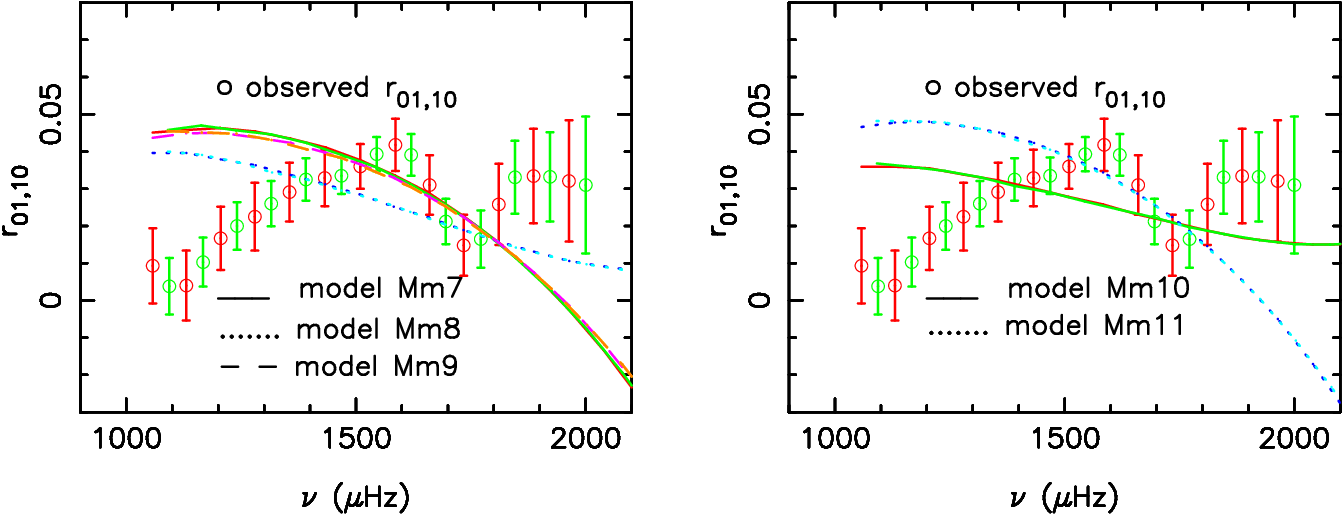}
\caption{The distributions of the observed and theoretical ratios \dra{} and \drb{}
as a function of frequency. The theoretical ratios are calculated from the corrected
frequencies $\nu_{corr}$ of models with the effective temperature and [Fe/H]
of \textbf{Molenda-Zakowicz et al. (2013)}. }
\label{fpgm1}
\end{figure}

Table \ref{tab2} lists the models with the effective temperature and [Fe/H] of \cite{mol13}.
The values of $\chi_{\nu}^{2}$ and $\chi_{\nu_{corr}}^{2}$
of the models are as large as those of models with the effective temperature and [Fe/H] of
\cite{brun12}. However, Figure \ref{fpgm1} shows that the ratios \dra{} and \drb{}
of the models are not in agreement with the observed ones. Our calculatios show
that the uncorrected ratios of these models also do not agree with the observed ones, i.e.,
the internal structures of these models do not match that of KIC 2837475.

The ratios of the models listed in Table \ref{tab2} decrease with the
increase in frequencies. \cite{liu14} showed that the ratios of
the modes whose inner turning points are located in overshooting
region of convective core can exhibit an increase behavior.
They also showed that the inner turning point, $r_{t}$, of the mode
with a frequency $\nu_{n,1}$ should be estimated by
\begin{equation}
\nu_{n,l}=f_{0}\frac{c(r_{t})}{r_{t}}\frac{\sqrt{l(l+1)}}{2\pi},
\label{tp}
\end{equation}
where $c(r_{t})$ is the adiabatic sound speed at radius $r_{t}$,
the value of the parameter $f_{0}$ is $2.0$.

\begin{figure}
\centering
\includegraphics[scale=0.35, angle=-90]{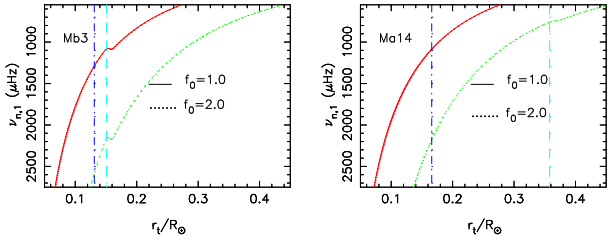}
\caption{The frequencies of models as a function of
inner turning point $r_{t}$ determined by Equation (\ref{tp}).
The vertical dash-dotted blue lines show the radius
of the convective core determined by Schwarzchild criterion.
The vertical dashed cyan lines
show the radius of the overshooting region. The solid red lines
show the frequencies determined by Equation (\ref{tp}) with $f_{0}=1.0$,
while the dotted green lines represent those determined by Equation (\ref{tp})
with $f_{0}=2.0$.}
\label{fpgtp}
\end{figure}

Figure \ref{fpgtp} shows the distributions of frequencies $\nu_{n,1}$ of models
as a function of inner turning point $r_{t}$ determined
by Equation (\ref{tp}). When the value of $f_{0}$ is equal to $2.0$,
the modes with $\nu_{n,1}$ less than $2200$ \dhz{} do not reach the overshooting
region of the convective core of model Mb3. The ratios \dra{} and \drb{}
of model Mb3 do not exhibit an increase behavior in the range between
1000 and 2200 \dhz{}. When the frequency is larger than 2200 \dhz{}, the ratios of
model Mb3 exhibit an obvious increase behavior (see Figure \ref{fpgb1}).
Thus the increase behavior of the ratios \dra{} and \drb{} may derive from
an effect of the overshooting of convective core.

The radius of overshooting region increase with the increase
in \dov{}, which leads to the fact that the increase behavior
can appear at low frequencies for the star with a large \dov{}.
Thus, we calculated the models with \dov{} as large as 1.8. With
the constraints of the effective temperature and [Fe/H] of \cite{brun12}
and \cite{mol13}, we did not obtain the models that can reproduce
the observed frequencies and ratios \dra{} and \drb{}.

\subsubsection{The Models with the Effective Temperature and [Fe/H] of Ammons}
The effective temperature determined by \cite{ammo06} is higher than that
estimated by \cite{mol13} but lower than the one determined by \cite{brun12}.
Moreover, the value of [Fe/H] estimated by \cite{ammo06} is higher than
those determined by \cite{brun12} and \cite{mol13}. We calculated models
with the effective temperature and [Fe/H] determined by \cite{ammo06}.
The value of \dov{} of the models with $\chi_{c}^{2}\leq1.0$ is
in the range of 0.0-0.4 and 1.0-1.6. The models with \dov{} between 0.0 and
0.4 can not reproduce the observed ratios \dra{} and \drb{}.
Table \ref{tab3} lists the models with the larger \dov{} and
a minimum $\chi_{\nu_{corr}}^{2}+\chi_{c}^{2}$ for a given mass and \dov{}.
The values of $\chi_{\nu_{corr}}^{2}$ of the models are as low as those of models
with the effective temperatures of \cite{brun12} or \cite{mol13}.
The value of $\chi_{\nu_{corr}}^{2}$ of the models with \dov{} = 1.4
is generally lower than that of other models. This indicates that the models
with \dov{} = 1.4 are better than other models.

\begin{table*}
\begin{center}
\renewcommand\arraystretch{1.0}
\caption[]{Parameters of Models with the Effective Temperature and [Fe/H] of \cite{ammo06}.
The symbols have the same meaning as those in Table \ref{tab2}.}
\label{tab3}
\begin{tabular}{p{0.70cm}cccccccccccccc}
  \hline\hline\noalign{\smallskip}
   Model & $M$ & \teff{}& $ L$ & $R$& age & $Z_{i}$ & $(Z/X)_{s}$ & $\alpha$
   & \dov{} & $X_{c}$  & $r_{cc}$ & $\chi_{\nu}^{2}$ & $\chi_{\nu_{corr}}^{2}$ & $\chi_{c}^{2}$ \\

  & (\dsm{}) & (K) & (\dsl{})  & (\dsr{}) & (Gyr) & &  & & & & (\dsr{}) & & \\
  \hline\hline\noalign{\smallskip}
 Ma1 & 1.44 & 6633 & 4.78 & 1.660 & 3.194 & 0.028 & 0.039 & 1.95 & 1.2 & 0.575 & 0.159 & 16.7 &  8.3 &  0.4 \\
 Ma2 & 1.44 & 6674 & 4.88 & 1.656 & 3.320 & 0.028 & 0.039 & 1.95 & 1.4 & 0.584 & 0.160 & 17.4 &  6.9 &  0.9 \\
 Ma3 & 1.44 & 6530 & 4.54 & 1.667 & 4.202 & 0.036 & 0.051 & 1.95 & 1.6 & 0.571 & 0.161 & 11.0 &  7.8 &  0.5 \\
\hline
 Ma4 & 1.46 & 6473 & 4.39 & 1.667 & 2.850 & 0.032 & 0.045 & 1.75 & 1.0 & 0.575 & 0.158 & 18.1 &  8.1 &  0.7 \\
 Ma5 & 1.46 & 6510 & 4.47 & 1.663 & 2.980 & 0.032 & 0.045 & 1.75 & 1.2 & 0.587 & 0.159 & 15.3 &  7.2 &  0.3 \\
 Ma6 & 1.46 & 6614 & 4.78 & 1.667 & 3.353 & 0.032 & 0.045 & 1.95 & 1.4 & 0.582 & 0.162 & 12.2 &  6.9 &  0.3 \\
 Ma7 & 1.46 & 6494 & 4.44 & 1.667 & 3.547 & 0.036 & 0.051 & 1.75 & 1.6 & 0.591 & 0.163 &  8.8 &  7.2 &  0.8 \\
\hline
 Ma8 & 1.48 & 6555 & 4.62 & 1.667 & 2.495 & 0.030 & 0.042 & 1.75 & 1.0 & 0.590 & 0.159 & 17.4 &  6.9 &  0.1 \\
 Ma9 & 1.48 & 6607 & 4.81 & 1.675 & 2.984 & 0.032 & 0.045 & 1.95 & 1.2 & 0.582 & 0.162 & 13.5 &  7.6 &  0.3 \\
 Ma10 & 1.48 & 6489 & 4.46 & 1.675 & 3.148 & 0.036 & 0.051 & 1.75 & 1.4 & 0.594 & 0.164 & 13.4 &  5.8 &  0.8 \\
 Ma11 & 1.48 & 6638 & 4.88 & 1.675 & 3.379 & 0.034 & 0.048 & 1.95 & 1.6 & 0.591 & 0.165 &  7.2 &  6.3 &  0.7 \\
\hline
 Ma12 & 1.50 & 6561 & 4.67 & 1.675 & 2.194 & 0.034 & 0.048 & 1.75 & 1.0 & 0.589 & 0.163 & 19.7 &  5.7 &  0.3 \\
 Ma13 & 1.50 & 6547 & 4.63 & 1.675 & 2.417 & 0.036 & 0.051 & 1.75 & 1.2 & 0.594 & 0.165 & 17.2 &  5.4 &  0.5 \\
 Ma14 & 1.50 & 6579 & 4.71 & 1.671 & 2.523 & 0.036 & 0.051 & 1.75 & 1.4 & 0.600 & 0.166 & 17.0 &  5.5 &  0.5 \\
 Ma15 & 1.50 & 6610 & 4.78 & 1.671 & 2.626 & 0.036 & 0.051 & 1.75 & 1.6 & 0.604 & 0.167 & 17.1 &  5.9 &  0.7 \\
\hline
 Ma16 & 1.52 & 6543 & 4.67 & 1.683 & 2.248 & 0.034 & 0.048 & 1.75 & 1.0 & 0.599 & 0.164 & 16.6 &  5.6 &  0.3 \\
 Ma17 & 1.52 & 6594 & 4.86 & 1.690 & 2.731 & 0.036 & 0.051 & 1.95 & 1.2 & 0.589 & 0.167 &  9.5 &  6.6 &  0.5 \\
 Ma18 & 1.52 & 6628 & 4.94 & 1.687 & 2.841 & 0.036 & 0.051 & 1.95 & 1.4 & 0.597 & 0.166 &  7.8 &  5.3 &  0.7 \\
 Ma19 & 1.52 & 6592 & 4.78 & 1.679 & 2.690 & 0.036 & 0.051 & 1.75 & 1.6 & 0.614 & 0.167 & 16.9 &  5.0 &  0.5 \\
\hline
 Ma20 & 1.54 & 6639 & 5.01 & 1.694 & 2.268 & 0.034 & 0.048 & 1.95 & 1.0 & 0.593 & 0.166 &  9.5 &  6.3 &  0.7 \\
 Ma21 & 1.54 & 6628 & 4.98 & 1.694 & 2.496 & 0.036 & 0.051 & 1.95 & 1.2 & 0.598 & 0.167 &  8.1 &  5.7 &  0.7 \\
 Ma22 & 1.54 & 6595 & 4.83 & 1.687 & 2.351 & 0.036 & 0.050 & 1.75 & 1.4 & 0.618 & 0.168 & 16.6 &  5.5 &  0.5 \\
 Ma23 & 1.54 & 6625 & 4.90 & 1.683 & 2.446 & 0.036 & 0.050 & 1.75 & 1.6 & 0.622 & 0.169 & 19.0 &  5.8 &  0.7 \\
\hline
 Ma24 & 1.56 & 6637 & 5.04 & 1.702 & 2.137 & 0.036 & 0.051 & 1.95 & 1.0 & 0.597 & 0.167 &  9.6 &  5.8 &  0.8 \\
 Ma25 & 1.56 & 6608 & 4.90 & 1.690 & 1.998 & 0.036 & 0.051 & 1.75 & 1.2 & 0.620 & 0.168 & 17.8 &  5.5 &  0.6 \\
 Ma26 & 1.56 & 6636 & 4.97 & 1.687 & 2.091 & 0.036 & 0.051 & 1.75 & 1.4 & 0.625 & 0.169 & 18.9 &  6.6 &  0.8 \\
 \noalign{\smallskip}\hline\hline
\end{tabular}
\end{center}
\end{table*}

\begin{figure*}
\centering
\includegraphics[scale=0.5, angle=-90]{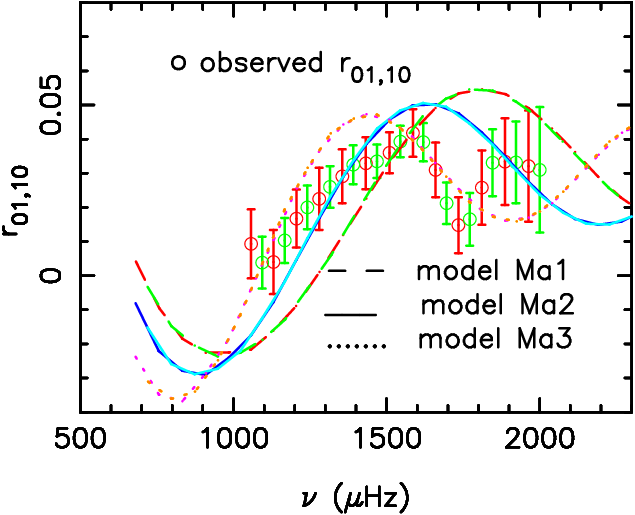}
\includegraphics[scale=0.5, angle=-90]{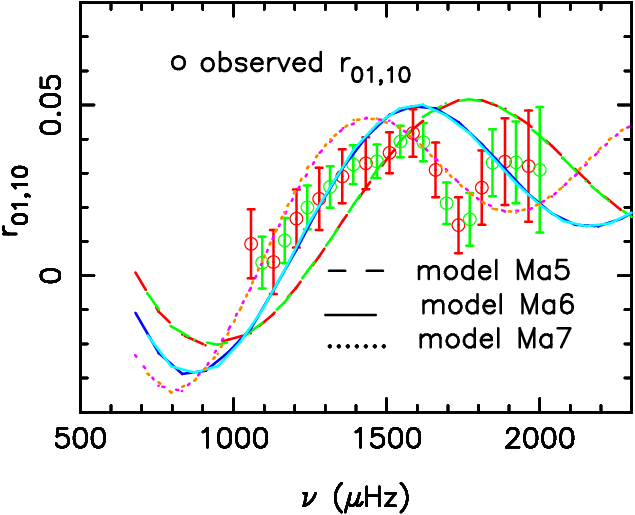}
\includegraphics[scale=0.5, angle=-90]{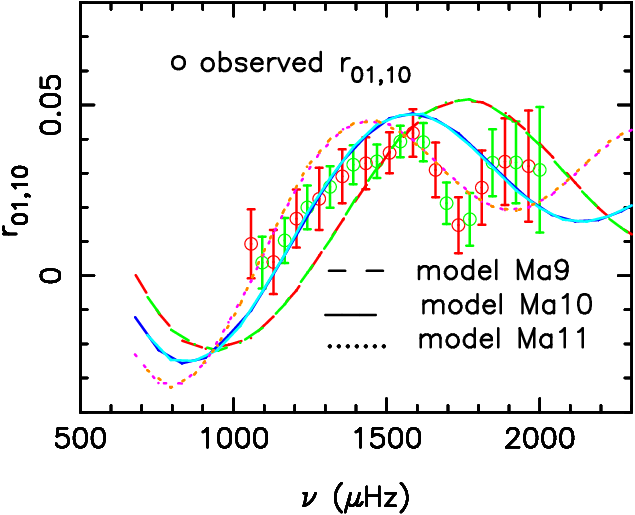}
\includegraphics[scale=0.5, angle=-90]{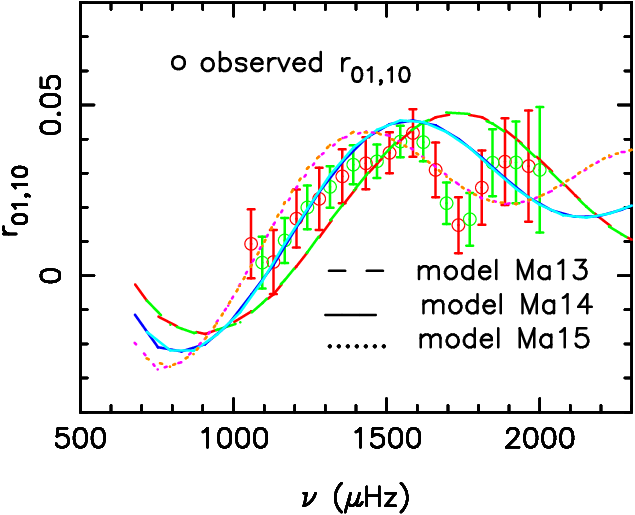}
\includegraphics[scale=0.5, angle=-90]{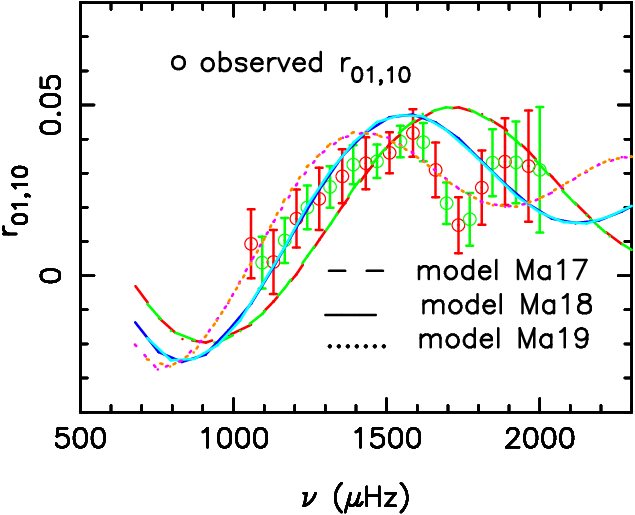}
\includegraphics[scale=0.5, angle=-90]{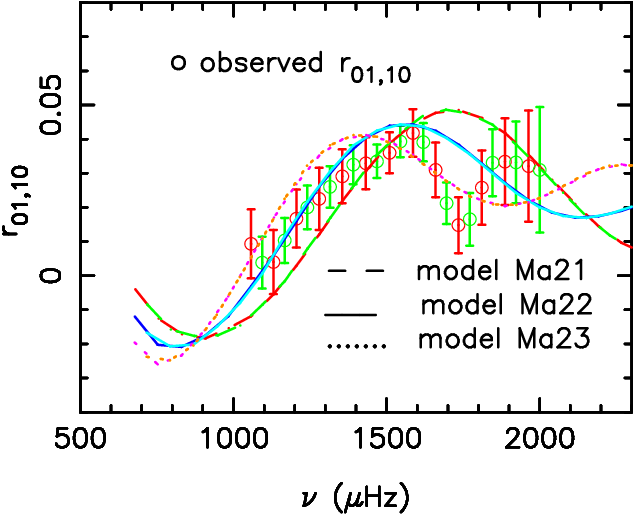}
\caption{The distributions of the observed and theoretical ratios \dra{} and \drb{}
as a function of frequency. The theoretical ratios are calculated from the corrected
frequencies $\nu_{corr}$ of models with the effective temperature and [Fe/H]
of \textbf{Ammons et al. (2006)}. }
\label{fpga1}
\end{figure*}

\begin{figure*}
\centering
\includegraphics[scale=0.5, angle=-90]{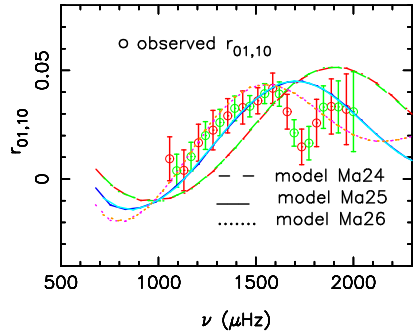}
\caption{The distributions of ratios \dra{} and \drb{} calculated from the corrected
frequencies of models Ma24, Ma25, and Ma26.}
\label{fpga2}
\end{figure*}

Figures \ref{fpga1} and \ref{fpga2} show the distributions of ratios \dra{} and \drb{}
of the models listed in Table \ref{tab3} as a function of frequency, which clearly show
that the observed ratios can be reproduced by the models with mass between about 1.46
and 1.56 \dsm{} and \dov{} between 1.2 and 1.6. This indicates that
KIC 2837475 may has a very thick overshooting region of convective core.

\begin{figure*}
\centering
\includegraphics[scale=0.5, angle=-90]{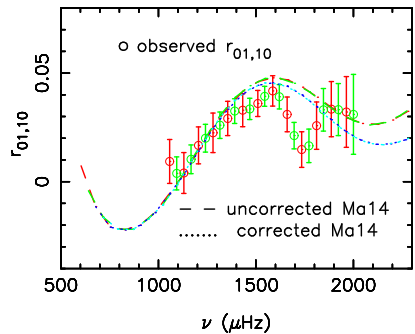}
\caption{The distribution of corrected and uncorrected ratios \dra{} and \drb{}
of model Ma14 as a function of frequency.}
\label{fpgun}
\end{figure*}

The distributions of ratios \dra{} and \drb{} can be affected by the correction
for the effect of near-surface layers. Thus we computed the ratios of uncorrected
frequencies of the models listed in Table \ref{tab3}. Figure \ref{fpgun} shows that
the difference between the uncorrected and corrected ratios mainly occurs at high
frequency. The distribution of the corrected ratios is similar to that
of uncorrected ones for these models. Therefore, the increase behavior of
the ratios of the models is uncorrelated with the effect of the correction.
It derives from an effect of the overshooting of convective core.
Even using the uncorrected ratios or the ratios of models with the minimum
$\chi_{\nu}^{2}$, our results are not changed.

\subsection{Fitting Curves of the Observed and Theoretical Ratios}

\cite{yang15} suggested using equation
\begin{equation}
r_{10}(\nu_{n,1})=\frac{2A\nu_{0}}{2\pi^{2}(\nu^{2}_{0}-\nu^{2}_{n,1})}
\sin(2\pi\frac{\nu_{n,1}}{\nu_{0}})+B_{0},
\label{nura}
\end{equation}
or
\begin{equation}
r_{10}(\nu_{n,1})=\frac{2A\nu_{n,1}}{2\pi^{2}(\nu^{2}_{0}-\nu^{2}_{n,1})}
\sin(2\pi\frac{\nu_{n,1}}{\nu_{0}})+B_{0},
\label{nurb}
\end{equation}
to describe the ratios \dra{} and \drb{} affected by the overshooting
of the convective core of stars, where the $A$ is a free parameter,
the $\nu_{0}$ is the frequency of the mode whose inner turning point
is located on the boundary between the radiative region
and the overshooting region of convective core, the $B_{0}$ is
the $r_{10}$ at $3\nu_{0}/2$. The value of $r_{10}(\nu_{n,1})$
decreases with frequency $\nu_{n,1}$ when the frequency is less
than $\nu_{0}$ but increases with frequency in the range of approximately
$\nu_{0}$ and $7\nu_{0}/4$, and has a maximum at about $7\nu_{0}/4$.
The distributions of ratios \dra{} and \drb{} of KIC 2837475 provide
a good opportunity for testing the formulae.

\begin{figure*}
\centering
\includegraphics[scale=0.5, angle=-90]{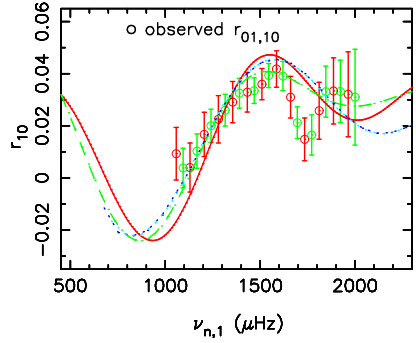}
\caption{The distribution of \drb{} as a function of $\nu_{n,1}$.
The dashed green line represents the ratio computed by
using equation (\ref{nura}) with $A=50\pi$, $\nu_{0}=900$ \dhz{}
and $B_{0}=0.032$. The solid red line shows the ratio calculated
by using equation (\ref{nurb}) with $A=50\pi$, $\nu_{0}=900$ \dhz{}
and $B_{0}=0.032$. The dotted lines represent the ratios of model Ma14. }
\label{fpgs}
\end{figure*}

The observed ratios of KIC 2837475 has a maximum at about $1580$ \dhz{}.
Thus the value of $\nu_{0}$ is approximately $900$ \dhz{}.
The value of the observed \drb{} of KIC 2837475
is about 0.032 at 1390 \dhz{}. \cite{yang15} showed that
the value of $A$ is $50\pi$ for KIC 11081729 and HD 49933.
With $A=50\pi$, $\nu_{0}=900$ \dhz{}, and $B_{0}=0.032$,
we calculated \drb{} for KIC 2837475 using Equations
(\ref{nura}) and (\ref{nurb}). Figure \ref{fpgs} represents the \drb{}
as a function of $\nu_{n,1}$. The \drb{} determined by Equation
(\ref{nura}) is more consistent with the observed ratios
than that determined by Equation (\ref{nurb}). However,
when the frequency $\nu_{n,1}$ is larger than $3\nu_{0}/2$,
the \drb{} determined by Equation (\ref{nurb}) is in better
agreement with the ratios of model Ma14 than that determined
by Equation (\ref{nura}). The observed and theoretical
ratios can be well reproduced by Equations (\ref{nura}) and
(\ref{nurb}). This further states that the increase behavior
in the ratios of KIC 2837475 arises from the effect of the
overshooting of convective core.

\subsection{Comparisons with Previous Models}
In the calculations, the effects of the overshooting of the convective core
were included that were not considered by \cite{metc14} in their models.
Moreover, the evolutions of models with a mass larger than $1.30$ \dsm{}
were computed without diffusion, while \cite{metc14}
considered the case with helium diffusion. The diffusion of helium and
heavy elements of \cite{thou94} could produce the almost metal-free and
pure-hydrogen models in the evolutions of models with a mass larger
than $1.30$ \dsm{} when the value of the mixing-length parameter is small.
For example, when the value of $\alpha$ is $1.75$, the value of the surface
metallicity of the model with $M=1.50$ and $Z_{i}=0.03$ can decrease to 0.003.
Thus we did not consider the diffusion in the evolutions of models with
a mass larger than 1.30 \dsm{}.

The mass of KIC 2837475 determined by \cite{metc14} is
$1.39\pm0.06$ \dsm{}, which is consistent with the mass
of the models with the effective temperature and [Fe/H] of \cite{mol13}.
However, the models cannot reproduce the observed ratios.
The mass of the models with the effective temperature and [Fe/H] of
\cite{ammo06} is obviously higher than that determined by \cite{metc14}.
The value of the large separation ($\Delta\nu$) and the frequency of
maximum seismic amplitude ($\nu_{max}$) is $76$ and $1522$ \dhz{}
\citep{appo12} for KIC 2837475, respectively. Using equation
\begin{equation}
M=(\frac{\Delta\nu}{\Delta\nu_{\odot}})^{-4}(\frac{\nu_{max}}{\nu_{max,\odot}})^{3}
(\frac{T_{\rm eff}}{T_{\rm eff,\odot}})^{3/2} M_{\odot},
\label{mas}
\end{equation}
we estimated the mass of KIC 2837475. The value of the mass is $1.493$ \dsm{}
for the effective temperature of $6562$ K of \cite{ammo06}, $1.554$ \dsm{}
for the effective temperature of $6740$ K of \cite{brun12}, and $1.460$ \dsm{}
for the effective temperature of $6462$ K of \cite{mol13}. Only the mass of
the models with the effective temperature and [Fe/H] of \cite{ammo06} is
consistent with that estimated by Equation (\ref{mas}).

\section{DISCUSSION AND SUMMARY}
\subsection{Discussion}
The values of the acoustic depth $\tau_{HeIIZ}$ and $\tau_{BCZ}$
\citep{mazu14} of model Ma14 are about 740 and 3770 $s$, respectively.
The changes caused by the glitch at the base of the convective envelope
or the second helium ionization zone have a periodicity of twice the
depth of the corresponding glitch \citep{mazu14}. For the model Ma14,
the periods are 1480 and 7540 $s$ which are corresponding to
the frequencies of 675.7 and 132.6 \dhz{}, respectively. The ratios \dra{} and
\drb{} of KIC 2837475 almost linearly increase in the range of approximately
1050 and 1600 \dhz{}. Thus the changes in the ratios could not derive
from the effects of the glitch at the base of the convective envelope
or the second helium ionization zone.

Figure \ref{fpgtp} shows that the modes with a frequency larger than about
760 \dhz{} penetrate into the overshooting region for model Ma14.
Figure \ref{fpgs} shows that an obvious increase behavior of ratios
of model Ma14 occurs at the frequencies larger than about 850 \dhz{}.
In addition, Figure \ref{fpga1} shows that the values of the frequencies of the
minimum and maximum \drb{} decrease with the increase in \dov{}.
The increase in \dov{} leads to the increase in the radius of
overshooting region, which results in the decrease in frequency
$\nu_{0}$, i.e., the location of the increase ratios moves to low
frequencies with the increase in \dov{}. Therefore, the increase behavior
of the ratios is related to the effects of overshooting of convective
core.

The minimum of the Equations (\ref{nura}) and (\ref{nurb}) does not
occur at $\nu_{0}$. For example, when the value of $\nu_{0}$ is 900 \dhz{},
the minimum of the Equation (\ref{nura}) occurs at 866 \dhz{};
but the minimum of the Equation (\ref{nurb}) occurs at 934 \dhz{}.
This leads to the difference between the results of Equations (\ref{nura})
and (\ref{nurb}) at $\nu_{n,1}<3\nu_{0}/2$. The minimum of the ratios
of model Ma14 is located at about 850 \dhz{}. Thus the results calculated
by Equation (\ref{nura}) is more consistent with theoretical ratios
when frequencies are less than $3\nu_{0}/2$.

The values of parameters $A$, $\nu_{0}$, and $B_{0}$ can be estimated
from the observed $\nu_{n,1}$ and \drb{} and are $34\pm10$ $\pi$, $900\pm35$ \dhz{},
and $0.030\pm0.002$ respectively for Equation (\ref{nura}), $23\pm7$ $\pi$,
$913\pm36$ \dhz{}, and $0.030\pm0.002$ respectively for Equation (\ref{nurb}).
The values of $\nu_{0}$ and $B_{0}$ determined by using the observed \drb{} and
Equation (\ref{nura}) are consistent with those determined by using the observed
\drb{} and Equation (\ref{nurb}). However, the values of $A$ have an obvious difference.
The Equations (\ref{nura}) and (\ref{nurb}) have a minimum
at about $\nu_{0}$ and a maximum at around $7\nu_{0}/4$.
The observed ratios of KIC 2837475 have not the minimum. Thus the observed ratios
are not suitable to determine the value of $A$.

Radiative levitation can lead to overabundances of Cr, Mn, Fe, and Ni
at the surface of stars with a mass larger than $1.30$ \dsm{} \citep{turc98}.
It is difficult to compute the radiative levitation effects in the evolutions
of a large sample of models because of numerical instabilities \citep{turc98}.
Thus the effects of radiative accelerations are not considered in our models
with a mass larger than $1.30$ \dsm{}.

Tables \ref{tab2} and \ref{tab3} clearly show that the value of the mass of
the obtained models is dependent on the effective temperature and [Fe/H]. Due to the
fact that the models with the effective temperature and [Fe/H] of \cite{brun12} or
\cite{mol13} can not reproduce the observed ratios \dra{} and \drb{}, only the models
with the effective temperature and [Fe/H] of \cite{ammo06} were used to estimate the
mass, radius, and age of KIC 2837475.

A large \dov{} means that an efficient mixing exists in
interior of stars. The mass of Procyon is $1.497\pm0.037$ \dsm{}.
The mass of our models for KIC 2837475 is in the range of \textbf{approximately}
1.46 and 1.56 \dsm{}. Moreover, there is the phenomenon of the double
or extended MS turnoffs in the color-magnitude diagram of
intermediate-age star clusters in the Large Magellanic
Cloud \citep{mack07, milo09, goud09}. The mass of the MS-turnoff
stars of the clusters is around 1.5 \dsm{} \citep{yang11b, yang13}.
\cite{yang13} showed that rapid rotation and extra mixing caused by
rotation can be used to explain the extended MS turnoffs. The surface
rotation period of KIC 2837475 is about 3.7 days \citep{garc14, mcq14}.
Rotation can lead to an increase in the convective core, which depends
on the efficiency of rotational mixing and rotation rate \citep{maed87, yang13}.
Furthermore, the extra mixing caused by rotation can mimic the
effect of the overshooting to a certain degree. This hints us
that the large \dov{} in KIC 2837475 may be related to rotation.
The large amount of overshoot may be the consequence of current inaccuracies
in the physical models that include many approximations.
The ``large \dov{}'' of KIC 2837475 may be not an exceptional
case in stars.

\subsection{Summary}

The calculations show that the observed ratios \dra{} and \drb{} of KIC 2837475
cannot be reproduced by the models with \dov{} $<1.2$. The value
of \dov{} is restricted between approximately 1.2 and 1.6 for KIC 2837475,
which is very close to that determined by \cite{guen14} for Procyon.
The effects of overshooting of the convective core can lead to an increase
behavior of ratios \dra{} and \drb{} of the p-modes whose
inner turning points are located in overshooting region.
For KIC 2837475, the inner turning points of p-modes with $l=1$
are located in the region of overshooting of the convective core.
The increase behaviors of the ratios \dra{} and \drb{}
of KIC 2837475 result from the effects of overshooting of the
convective core. KIC 2837475 shows the existence of large \dov{}
and also confirms that the increase behavior of ratios \dra{} and \drb{}
derives from the direct effects of overshooting of
the convective core.

The increase in \dov{} can lead to the increase in the radius of overshooting
region, which results in the fact that the low-frequency modes can penetrate
into the overshooting region. the ratios \dra{} and \drb{} of p-modes whose
inner turning points are located in the overshooting region can exhibit an
increase behavior. Thus the distributions of the ratios \dra{} and \drb{} are
sensitive to \dov{}. The distributions of the observed and theoretical ratios
\dra{} and \drb{} of KIC 2837475 are reproduced well by the equations
of \cite{yang15}, which also indicates that the increase behavior of the ratios
arises from the effects of overshooting of the convective core.
For the evolutions without the effects of diffusion, with the constraints
of \cite{ammo06} spectroscopic results and the observed \dra{} and \drb{},
the observational constraints favor a star with $M=1.490\pm0.018$ \dsm{},
$R=1.67\pm0.01$ \dsr{}, age $=2.8\pm0.4$ Gyr, and $1.2\lesssim$ \dov{}
$\lesssim1.6$ for KIC 2837475, where the uncertainty indicates
the 68\% level confidence interval determined by probability distribution
function.

\section*{Acknowledgments}
We thank the anonymous referee for enlightening comments and
acknowledge the support from the NSFC 11273012, 11273007,
the Fundamental Research Funds for the Central Universities,
and the HSCC of Beijing Normal University.

\end{document}